# Hard Decision Cooperative Spectrum Sensing Based on Estimating the Noise Uncertainty Factor

Hossam M. Farag
Dept. of Electrical Engineering, Aswan University, Egypt.
hossam.farag@aswu.edu.eg

Ehab Mahmoud Mohamed
Current: Dept. of Information & Communication Technology, Osaka University, Japan.
Permanent: Dept. of Electrical Engineering, Aswan University, Egypt.
ehab@wireless.comm.eng.osaka-u.ac.jp

*Abstract*—Spectrum Sensing (SS) is one of the most challenging issues in Cognitive Radio (CR) systems. Cooperative Spectrum Sensing (CSS) is proposed to enhance the detection reliability of a Primary User (PU) in fading environments. In this paper, we propose a hard decision based CSS algorithm using energy detection with taking into account the noise uncertainty effect. In the proposed algorithm, two dynamic thresholds are toggled based on predicting the current PU activity, which can be successfully expected using a simple successive averaging process with time. Also, their values are evaluated using an estimated value of the noise uncertainty factor. These dynamic thresholds are used to compensate the noise uncertainty effect and increase (decrease) the probability of detection (false alarm), respectively. Theoretical analysis is performed on the proposed algorithm to deduce its enhanced false alarm and detection probabilities compared to the conventional hard decision CSS. Moreover, simulation analysis is used to confirm the theoretical claims and prove the high performance of the proposed scheme compared to the conventional CSS using different fusion rules.

## I. INTRODUCTION

Cognitive Radio (CR) technique has been proposed to resolve the conflicts between spectrum scarcity and spectrum underutilization [1]. It allows Cognitive Radio users (CRs) to share the spectrum with a Primary User (PU) by opportunistic accessing. Spectrum Sensing (SS) is the most critical issue in CR technology, since it needs to detect the presence and the absence of the PU accurately and swiftly; high efficient SS technique results in highly operated CR node. Common methods of SS are Energy Detection (ED), Cyclostationary Detection (CD) and Matched Filter Detection (MFD) [2]. ED is the most preferable approach for SS due to its simplicity and applicability as it does not need any prior knowledge about the PU signal. In some specific environments, multi-path fading, shadowing and hidden node problems may cause the disability of CRs to accurately detect the presence of PUs. In order to solve such problems, Cooperative Spectrum Sensing (CSS) is introduced to achieve an improved sensing performance [3].

In ED based CSS, each CR individually measures the energy received from the PU, and the final decision is made based on the Fusion Center (FC) rule. In hard decision CSS [4], each CR individually decides the current PU activity based on its own measurements. Then, it sends its final binary decision (presence/absence) to the FC, which in turn applies the AND, OR, or Majority rule on the collected CRs decisions to arrive at a final decision about the current PU activity. On the other hand, in soft decision CSS, the CRs work as energy sensors for the FC. They frequently send their own PU energy measurements to the FC.

Then, the FC applies one of the energy combining schemes on the collected energy measurements, such as Square-Law Combining (SLC), Square-Law Selection (SLS) and Maximum Ratio Combining (MRC) to arrive at a final decision about the current PU activity [5]. Although soft decision CSS highly outperforms hard decision CSS, it requires higher communication bandwidth and higher computational complexity, which increases the total cost of the Cognitive Radio Networks (CRNs) [2]. In this paper, we will focus on the low bandwidth/low complexity hard decision CSS through enhancing its spectrum sensing performance.

ED is highly affected by a well-known phenomenon called the SNR wall that prohibits the PU detection to be robust [6]. A lot of research work have been done to overcome the noise uncertainty effect in CRNs to reduce the total number of CRs required for efficient CSS [7]-[11]. In [7], [8], eigenvalue based SS is proposed as an alternative to the conventional ED based SS to enhance the sensing performance of the CRNs particularly in the presence of noise uncertainty. Although eigenvalue based SS outperforms the ED based SS, it suffers from high computational complexity comes from covariance matrix construction and eigenvalue calculations. Another existing approach, proposed in [9]-[11], is based on using different threshold levels to enhance the ED performance against noise uncertainty with marginal increase in computational complexity, which is our main concern in this paper. Although the researchers agreed that the threshold levels should be dynamically changed based on the noise uncertainty values, they did not give a practical methodology of how we can estimate these values. At the best of our knowledge, most of the researchers assumed that the noise uncertainty factor is a given parameter which is well known to the ED beforehand. In addition, there is a lack of an efficient criterion based on which the dynamic thresholds should be toggled in a way to increase the probability of detection ($P_d$) and decrease the probability of false alarm ($P_{fa}$).

In this paper, to design a low complexity ED based hard decision CSS algorithm while overcoming the noise uncertainty effect, a practical methodology for estimating the noise uncertainty factor is proposed. In which, each CR evaluates its noise uncertainty factor by considering its noise variance history over a predefined period. To efficiently overcome the noise uncertainty effect, two dynamic thresholds are utilized by each CR evaluated based on the pre-estimated noise uncertainty factor. A simple criterion is also proposed to switch between the dynamic thresholds to increase the $P_d$ and decrease the $P_{fa}$. This criterion is based on predicting the current PU activity, in which each CR node predicts, to some extent, the presence or the absence of the PU in the current sensing event by observing its recent activities. This can be done with a high probability thanks to fact that the time required by a PU to change its status (ON/OFF) is negligible to the time a PU remains at a certain status (ON/OFF). By predicting the current PU activity, the two dynamic thresholds are toggled to maximize $P_d$, if the PU is predicted to be present or to minimize the $P_{fa}$ if the PU is predicted to be absent by alleviating the noise uncertainty effect.

Theoretical analysis is performed on the proposed algorithm to derive its improved $P_d$ and $P_{fa}$ over the conventional ones. Moreover, simulation analysis is conducted to confirm the theoretical claims and prove the high efficiency of proposed scheme compared to the conventional one, in which noise uncertainty effect is not considered, using the same number of CRs. In addition, it is shown that a lower number of CRs are required by the proposed scheme to obtain the same performance like the conventional one.

The rest of this paper is organized as follows. Section II reviews the conventional hard decision CSS. Section III describes the proposed hard decision CSS algorithm. Section IV provides the theoretical analysis of the proposed algorithm. Simulation analysis is given in Sect. V. Complexity analysis is investigated in Sect. VI, followed by the conclusion in Sect. VII.

## II. THE CONVENTIONAL HARD DECISION CSS

The CRN considered in this paper consists of $K$ CR nodes and a secondary base station (BS) including the FC, which are located in the area of a primary base station (PU). For the sake of simplicity, noiseless reporting channels are assumed between the CRs and the secondary BS.

The received signal at the *j-th* CR, $1 \leq j \leq K$, from the PU, can be defined as:

$$y(n)_j = \begin{cases} w(n)_j & H_0 \\ h_j s(n) + w(n)_j & H_1 \end{cases}, \quad (1)$$

$n = 1, 2, \dots, N$, where $N$ is the total number of received samples, $y(n)_j$ is the received signal at the *j-th* CR, and $w(n)_j$ is the Additive White Gaussian Noise (AWGN) at the *j-th* CR. $s(n)$ is the PU signal, and $h_j$ is the complex channel response between the PU and the *j-th* CR; $h_j$ will be assumed as Rayleigh flat fading channel. Hypothesis $H_0$ states that PU is absent, and hypothesis $H_1$ states that PU is present.

Using ED, the received energy $E_j$ at the *j-th* CR is given as:

$$E_j = \sum_{n=1}^{N} y(n)_j^2. \quad (2)$$

For sufficiently large $N$ ($N \gg 1$), according to the Central Limit Theorem (CLT), the received energy $E_j$ can be approximated as Gaussian [12]:

$$E_j \sim \begin{cases} \mathcal{N}(N\sigma^2, 2N\sigma^4) & H_0 \\ \mathcal{N}\left(N\sigma^2(\gamma_j+1), 2N\sigma^4(\gamma_j+1)^2\right) & H_1 \end{cases}, \quad (3)$$

where $\gamma_j$ is the signal-to-noise ratio (SNR) of the *j-th* CR and $\sigma^2$ is the noise variance. With Rayleigh fading channel between the PU and the CRs, the false alarm probability $P_{fa,j}$ and the detection probability $\overline{P_{d,j}}$ for the *j-th* CR can be expressed as [12]:

$$P_{fa,j} = P_r(E_j \geq \lambda | H_0) = Q\left(\frac{\lambda - N\sigma^2}{\sqrt{2N}\sigma^2}\right), \quad (4)$$

$$\overline{P_{d,j}} = P_r(E_j \geq \lambda | H_1) = \int_0^\infty P_{d,j} f(\gamma_j) d\gamma_j, \quad (5)$$

where $Q(.)$ is the standard Gaussian complementary cumulative distribution function, $\lambda$ is the decision threshold, $P_{d,j}$ is the detection probability of the *j-th* CR under AWGN channel, and $f(\gamma_j)$ is the probability density function (PDF) of $\gamma_j$ under Rayleigh fading channel. $P_{d,j}$ and $f(\gamma_j)$ are expressed as [12]:

$$P_{d,j} = Q\left(\frac{\lambda - N\sigma^2(\gamma_j+1)}{\sqrt{2N}\sigma^2(\gamma_j+1)}\right), \quad (6)$$

$$f(\gamma_j) = \frac{1}{\overline{\gamma}} e^{-\frac{\gamma_j}{\overline{\gamma}}}, \quad \gamma_j \geq 0, \quad (7)$$

where $\overline{\gamma}$ is the average SNR.

CSS takes the advantage of spatial diversity to improve the SS performance against fading, shadowing, and hidden PU problems. For hard decision CSS, each CR makes its own binary decision about the presence or the absence of the PU, then it sends a one bit information '1' (PU exists) or '0' (PU does not exist), respectively to the FC. Decision rule such as AND, OR or Majority rule is applied on the collected binary data by the FC to make a final decision about the PU existence.

Using AND rule, the FC decides that a PU is present if all CRs decide that the PU is present. The detection probability $P_{d,AND}$ and the false alarm probability $P_{fa,AND}$ using the AND rule are given as [4]:

$$P_{d,AND} = \left(\overline{P_{d,j}}\right)^K, \quad (8)$$

$$P_{fa,AND} = \left(P_{fa,j}\right)^K. \quad (9)$$

Using OR rule, the FC decides that a PU is present if only one of the CRs decides that the PU is present. The detection probability $P_{d,OR}$ and the false alarm probability $P_{fa,OR}$ using the OR rule are defined as [4]:

$$P_{d,OR} = 1 - \left(1 - \overline{P_{d,j}}\right)^K, \quad (10)$$

$$P_{fa,OR} = 1 - \left(1 - P_{fa,j}\right)^K. \quad (11)$$

Using Majority rule, the FC decides that a PU is present if at least $l$ out of $K$ CRs decide that the PU is present. The detection probability $P_{d,Maj}$ and the false alarm probability $P_{fa,Maj}$ of the Majority rule are defined as follows [4]:

$$P_{d,Maj} = \sum_{t=l}^{K} \binom{K}{t} \left(\overline{P_{d,j}}\right)^t \left(1 - \overline{P_{d,j}}\right)^{K-t}, \quad (12)$$

$$P_{fa,Maj} = \sum_{t=l}^{K} \binom{K}{t} \left(P_{fa,j}\right)^t \left(1 - P_{fa,j}\right)^{K-t}. \quad (13)$$

In the conventional hard decision CSS, a fixed decision threshold $\lambda$ is used by the CRs in the PU detection operation. Consequently, the ED done by the CRs will be bounded by the SNR wall due to noise uncertainty effect [6]. As a result, to obtain a sufficient sensing performance, a high number of CRs should be used in the CSS operation.

## III. THE PROPOSED HARD DECISION CSS

In this section, we introduce the proposed hard decision CSS scheme using AND, OR, and Majority fusion rules. In this scheme, two dynamic thresholds are used by each CR to alleviate the noise uncertainty in their PU energy measurements. These thresholds are calculated based on estimating the noise uncertainty factor and toggled based on an expectation of the current PU activity. Towards that, each CR node individually constitutes the PU activity profile during a predefined period ($L$) by storing the $L$-1 consecutive received PU energy values $E_j(i)$, $1 \leq i \leq L$. In addition, each CR estimates and stores the noise variances of these $L$-1 received PU signals $\sigma_j^2(i)$. Based on these stored PU energy values, each CR can predict, to some extent, the current PU status by evaluating the average value of the $L$ consecutive energy measurements $E_{avg}(j)$.

The averaging process includes the measured PU energy of the current observation period $E_j(L)$, of which the final decision about the PU presence/absence should be taken, as follows:

$$E_{avg}(j) = \frac{1}{L}\sum_{i=1}^{L} E_j(i) \qquad 1 \leq j \leq K. \qquad (14)$$

Concurrently, the noise uncertainty factor $\rho(j)$ in the $j$-th CR energy measurements is estimated by the $j$-th CR based on the stored $L$-1 noise variances and the current estimated noise variance $\sigma_j^2(L)$, as follows:

$$\rho(j) = \frac{\max_{1 \leq i \leq L} \sigma_j^2(i)}{\frac{1}{L}\sum_{i=1}^{L} \sigma_j^2(i)}. \qquad (15)$$

After calculating $E_{avg}(j)$ and $\rho(j)$, the current status of the PU can be predicted by each CR by comparing $E_{avg}(j)$ with a predefined threshold $\lambda$ which is calculated based on the Constant False Alarm Rate (CFAR) criterion by inverting (4). That is, if $E_{avg}(j) \geq \lambda$, the CR expects to some extent that the PU is currently present; thanks to the fact that the time required by a PU to change its status (ON/OFF) is negligible to the time a PU remains at certain status (ON/OFF). Consequently, a new decision threshold at the $j$-th CR $\lambda_{new}(j)$ is set to equal $\lambda/\rho(j)$, which in turn results in maximizing the probability of detection, as it combats the noise uncertainty effect $\rho(j)$ at each CR. On the other hand, if $E_{avg} < \lambda$, the CR expects, to some extent, that the PU is currently absent, so $\lambda_{new}(j)$ is set to equal $\rho(j)\lambda$ to minimize the probability of false alarm by alleviating the noise uncertainty effect at each CR. By comparing $E_j(L)$ with $\lambda_{new}(j)$, a local binary decision is made by a $j$-th CR as follows:

$$E_j(L) \geq \lambda_{new}(j) \qquad PU\ is\ present,$$
$$E_j(L) < \lambda_{new}(j) \qquad PU\ is\ absent. \qquad (16)$$

Then, each CR node sends its own binary decision to the FC, which in turn applies AND, OR, or Majority rule on the collected binary decisions to make a final decision about the existence of the PU in the current observation period $L$.

## IV. THEORETICAL ANALYSIS OF THE PROPOSED SCHEME

In this section, the enhanced false alarm and detection probabilities of the proposed scheme over the conventional ones are derived. The stored energy values $E_j(i)$ are assumed to be normally distributed (as given in (3)) and mutually independent (as they represent the energy of the sensed signal at time instants separated by multiples of the sensing period $N$). Since $E_{avg}(j)$ is the average over independent and identically distributed (i.i.d.) Gaussian random variables, it is also considered to be normally distributed:

$$E_{avg}(j) \sim \mathcal{N}\left(\mu_{avg}(j), \sigma_{avg}^2(j)\right), \qquad (17)$$

where $\mu_{avg}(j)$ and $\sigma_{avg}^2(j)$ are expressed as follows:

$$\mu_{avg}(j) = \frac{M}{L} N\sigma^2(\gamma_j + 1) + \frac{L-M}{L} N\sigma^2, \qquad (18)$$

$$\sigma_{avg}^2(j) = \frac{M}{L^2} 2N\sigma^4(\gamma_j + 1)^2 + \frac{L-M}{L^2} 2N\sigma^4. \qquad (19)$$

where $M \in \{0,1,2,...,L\}$ is the number of sensing events where the PU is actually present. Based on the proposed algorithm, the false alarm probability $P_{fa,j}^{Prop}$ and the detection probability $P_{d,j}^{Prop}$ of the $j$-th CR under AWGN channel can be given as:

$$P_{fa,j}^{Prop} = \{P_r\left(E_{avg}(j) \geq \lambda, E_j(L) \geq \lambda/\rho(j)\right)\}_{H_0}$$
$$+ \{P_r(E_{avg}(j) < \lambda, E_j(L) \geq \rho(j)\lambda)\}_{H_0},$$
$$= \{P_r(E_j(L) \geq \lambda/\rho(j)|E_{avg}(j) \geq \lambda). P_r(E_{avg}(j) \geq \lambda)\}_{H_0}$$
$$+ \{P_r(E_j(L) \geq \rho(j)\lambda|E_{avg}(j) < \lambda). P_r(E_{avg}(j) < \lambda)\}_{H_0}, \quad (20)$$

$$P_{d,j}^{Prop} = \{P_r\left(E_{avg}(j) \geq \lambda, E_j(L) \geq \lambda/\rho(j)\right)\}_{H_1}$$
$$+ \{P_r(E_{avg}(j) < \lambda, E_j(L) \geq \rho(j)\lambda)\}_{H_1},$$
$$= \{P_r(E_j(L) \geq \lambda/\rho(j)|E_{avg}(j) \geq \lambda). P_r(E_{avg}(j) \geq \lambda)\}_{H_1}$$
$$+ \{P_r(E_j(L) \geq \rho(j)\lambda|E_{avg}(j) < \lambda). P_r(E_{avg}(j) < \lambda)\}_{H_1}. \quad (21)$$

$E_{avg}(j)$ is calculated over a representative number of $L$ stored energy values $E_j(i)$ to accurately predict the current activity of the PU. Since the average of a relatively large set of values is not significantly affected, in general, by the value of a single element, it is reasonable, for $L$ sufficiently large, to assume that $E_{avg}(j)$ is approximately independent of $E_j(i)$ regardless of the actual channel status. Consequently, we get:

$$P_r(E_j(L) \geq \lambda/\rho(j)|E_{avg}(j) \geq \lambda) \approx P_r\left(E_j(L) \geq \lambda/\rho(j)\right), \quad (22)$$
$$P_r(E_j(L) \geq \rho(j)\lambda|E_{avg}(j) < \lambda) \approx P_r(E_j(L) \geq \rho(j)\lambda). \quad (23)$$

Using (22) and (23) we get:

$$P_{fa,j}^{Prop} \approx \{P_r(E_{avg}(j) \geq \lambda). P_r\left(E_j(L) \geq \lambda/\rho(j)\right)\}_{H_0}$$
$$+ \{P_r(E_{avg}(j) < \lambda). P_r(E_j(L) \geq \rho(j)\lambda)\}_{H_0}, \quad (24)$$

$$P_{d,j}^{Prop} \approx \{P_r(E_{avg}(j) \geq \lambda). P_r\left(E_j(L) \geq \lambda/\rho(j)\right)\}_{H_1}$$
$$+ \{P_r(E_{avg}(j) < \lambda). P_r(E_j(L) \geq \rho(j)\lambda)\}_{H_1}. \quad (25)$$

Based on (17) and (24), $P_{fa,j}^{Prop}$ can be given as:

$$P_{fa,j}^{Prop} \approx Q\left(\frac{\lambda - \mu_{avg}(j)}{\sigma_{avg}(j)}\right). Q_{fa,j}(\lambda/\rho(j))$$
$$+ \left[1 - Q\left(\frac{\lambda - \mu_{avg}(j)}{\sigma_{avg}(j)}\right)\right]. Q_{fa,j}(\rho(j)\lambda),$$
$$\approx Q\left(\frac{\lambda - \mu_{avg}(j)}{\sigma_{avg}(j)}\right). [Q_{fa,j}(\lambda/\rho(j)) - Q_{fa,j}(\rho(j)\lambda)] + Q_{fa,j}(\rho(j)\lambda),$$
$$(26)$$

where:

$$Q_{fa,j}(\lambda/\rho(j)) = Q\left(\frac{\lambda/\rho(j) - N\sigma^2}{\sqrt{2N}\sigma^2}\right), \qquad (27)$$

$$Q_{fa,j}(\rho(j)\lambda) = Q\left(\frac{\rho(j)\lambda - N\sigma^2}{\sqrt{2N}\sigma^2}\right). \qquad (28)$$

By analogy, $P_{d,j}^{Prop}$ can be written as:

$$P_{d,j}^{Prop} \approx Q\left(\frac{\lambda - \mu_{avg}(j)}{\sigma_{avg}(j)}\right). Q_{d,j}(\lambda/\rho(j))$$
$$+ \left[1 - Q\left(\frac{\lambda - \mu_{avg}(j)}{\sigma_{avg}(j)}\right)\right]. Q_{d,j}(\rho(j)\lambda),$$
$$\approx Q\left(\frac{\lambda - \mu_{avg}(j)}{\sigma_{avg}(j)}\right). [Q_{d,j}(\lambda/\rho(j)) - Q_{d,j}(\rho(j)\lambda)] + Q_{d,j}(\rho(j)\lambda),$$
$$(29)$$

where:

$$Q_{d,j}(\lambda/\rho(j)) = Q\left(\frac{\lambda/\rho(j) - N\sigma^2(\gamma_j + 1)}{\sqrt{2N}\sigma^2(\gamma_j + 1)}\right), \quad (30)$$

$$Q_{d,j}(\rho(j)\lambda) = Q\left(\frac{\rho(j)\lambda - N\sigma^2(\gamma_j + 1)}{\sqrt{2N}\sigma^2(\gamma_j + 1)}\right). \quad (31)$$

Using $P_{d,j}^{Prop}$, the detection probability of the proposed algorithm $\overline{P_{d,j}^{Prop}}$ under Rayleigh fading channel at the *j-th* CR is given as:

$$\overline{P_{d,j}^{Prop}} = \int_0^\infty P_{d,j}^{Prop} f(\gamma_j) d\gamma_j. \quad (32)$$

The overall false alarm and detection probabilities of the proposed scheme using AND fusion rule $P_{fa,AND}^{Prop}$ and $P_{d,AND}^{Prop}$ become:

$$P_{fa,AND}^{Prop} = \left(P_{fa,j}^{Prop}\right)^K, \quad (33)$$

$$P_{d,AND}^{Prop} = \left(\overline{P_{d,j}^{Prop}}\right)^K. \quad (34)$$

Also, the overall false alarm and detection probabilities using OR rule $P_{fa,OR}^{Prop}$ and $P_{d,OR}^{Prop}$ become:

$$P_{fa,OR}^{Prop} = 1 - \left(1 - P_{fa,j}^{Prop}\right)^K, \quad (35)$$

$$P_{d,OR}^{Prop} = 1 - \left(1 - \overline{P_{d,j}^{Prop}}\right)^K. \quad (36)$$

Also, using Majority fusion rule, $P_{fa,Maj}^{Prop}$ and $P_{d,Maj}^{Prop}$ become:

$$P_{fa,Maj}^{Prop} = \sum_{t=l}^{K} \binom{K}{t} \left(P_{fa,j}^{Prop}\right)^t \left(1 - P_{fa,j}^{Prop}\right)^{K-t}, \quad (37)$$

$$P_{d,Maj}^{Prop} = \sum_{t=l}^{K} \binom{K}{t} \left(\overline{P_{d,j}^{Prop}}\right)^t \left(1 - \overline{P_{d,j}^{Prop}}\right)^{K-t}. \quad (38)$$

Given that the PU is actually absent ($H_0$) during the sensing period $L$, a reliable PU activity predictor will get $Q\left(\frac{\lambda - \mu_{avg}(j)}{\sigma_{avg}(j)}\right) \approx 0$. With such a value, from (26), we get $P_{fa,j}^{Prop} \approx Q_{fa,j}(\rho(j)\lambda)$, which is lower than the conventional $P_{fa,j}$ given in (4). Thus, $P_{fa,AND}^{Prop}$, $P_{fa,OR}^{Prop}$, and $P_{fa,Maj}^{Prop}$ in (33), (35), and (37) will be lower than conventional $P_{fa,AND}$, $P_{fa,OR}$, and $P_{fa,Maj}$ given in (9), (11), and (13), respectively. By analogy, given $H_1$, the reliable predictor gives $Q\left(\frac{\lambda - \mu_{avg}(j)}{\sigma_{avg}(j)}\right) \approx 1$, so from (29), we get $P_{d,j}^{Prop} \approx Q_{d,j}(\lambda/\rho(j))$, which is greater than $P_{d,j}$ given in (6). Therefore, $P_{d,AND}^{Prop}$, $P_{d,OR}^{Prop}$, and $P_{d,Maj}^{Prop}$ in (34), (36), and (38) will be higher than conventional $P_{d,AND}$, $P_{d,OR}$, and $P_{d,Maj}$ given in (8), (10), and (12), respectively.

The PU activity predictor will inaccurately predict the current PU status at the time when the PU changes its status from ON to OFF and vice versa. This will increase $P_{fa,j}^{Prop}$ and decrease $P_{d,j}^{Prop}$ compared to the conventional $P_{fa,j}$ and $P_{d,j}$, respectively. But, these cases will not affect the overall improved performance of the proposed scheme because the time required by a typical PU to change its status (ON/OFF) is negligible to the time the PU remains at a certain status (ON/OFF). Therefore, using the same number of CRs, it is expected that the proposed hard decision CSS outperforms the conventional one.

## V. SIMULATION ANALYSIS

In the conducted simulations, a real communication environment is simulated for the CRN using BPSK signal and Rayleigh flat fading channels between the PU and all CRs. In addition, noiseless reporting channels are assumed between the FC and the CRs. Monte Carlo computer simulations are used to prove the effectiveness of the proposed scheme using Receiver Operating Characteristics (ROC) curves. In which, we draw the relationship between $P_d$ and $P_{fa}$.

First, we optimize the performance of the proposed scheme against its critical parameters such as the value of the consecutive energy records $L$ and the number of used CRs using OR fusion rule. Figure 1 shows the ROC curves using different $L$ values, i.e., $L = 5$, 10, 15, and 20, and SNR of -20 dB. The number of used samples $N$ is equal to 1000 samples and the number of CRs $K$ is equal to 3. From this figure, as we increase the number of stored records $L$, the performance of the proposed scheme is enhanced. Based on the trade-off between performance and complexity, we choose $L = 15$ as a sufficient value for $L$. In Fig. 2, the number of CRs $K$ is adjusted using OR fusion rule, the adjusted value of $L = 15$, $N = 1000$ and SNR = -20 dB. From this figure, $K = 7$ is chosen as a sufficient number of CRs.

Figure 3 shows performance comparisons between the proposed hard decision CSS and the conventional hard decision CSS using AND, OR, and Majority fusion rules. To explicitly show the high impact of the proposed hard decision CSS, we also give the performance of the best conventional CSS, which is the conventional soft decision CSS using MRC [5]. In this simulation, we use $L = 15$, $K = 7$, $N = 1000$, $l = 3$ for the Majority fusion rule, and SNR = -15 dB. From this figure, at $P_{fa} = 0.1$ (which is a rated parameter for typical CRN design), the proposed algorithm increases $P_d$ by about 3 times more than the $P_d$ provided by the conventional hard decision CSS using different fusion rules (AND, OR, and Majority rules).

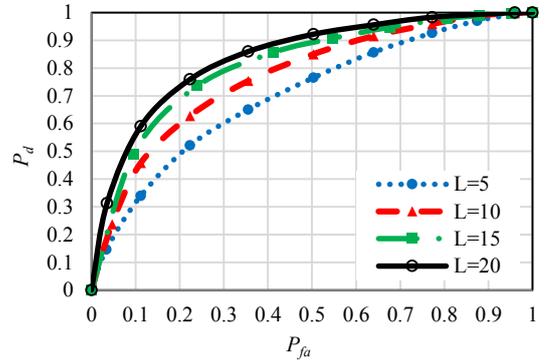

Fig. 1. ROC curves using different values of *L*.

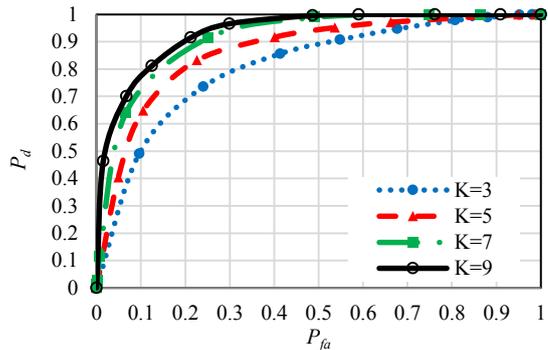

Fig. 2. ROC curves using different values of *K*.

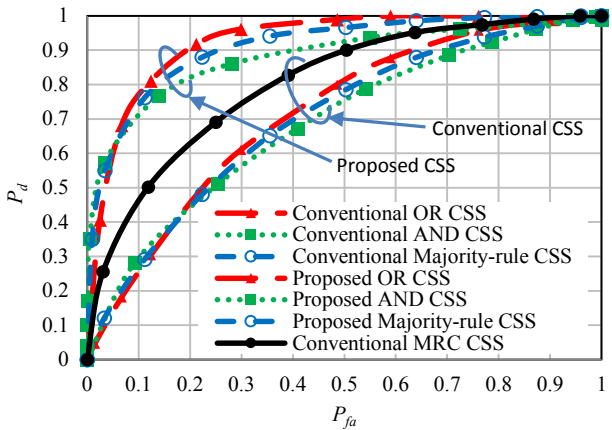

Fig. 3. Performance comparisons.

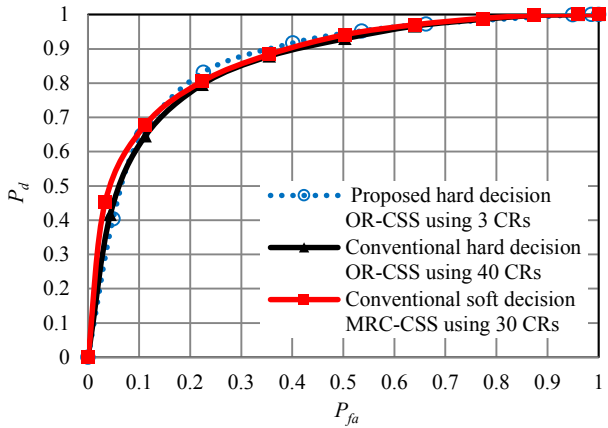

Fig. 4. Complexity comparisons.

Furthermore, it increases $P_d$ by about 1.5 times more than the $P_d$ provided by the conventional soft decision CSS using MRC, which is the best conventional CSS scheme [5].

The highly enhanced performance of the proposed hard decision CSS comes from alleviating the noise uncertainty effect in the CRs energy measurements, of which function is not provided by the conventional CSS. As a result, the proposed hard decision CSS outperforms not only the conventional hard decision CSS but also the conventional soft decision CSS using MRC.

## VI. COMPLEXITY ANALYSIS

Figure 4 shows complexity comparisons between the proposed hard decision CSS using OR fusion rule, the conventional hard decision CSS using OR fusion rule and the conventional soft decision CSS using MRC. In this simulation, we use $L = 15$, $N = 1000$, and SNR = -15 dB. It is clearly shown that the proposed scheme only using 3 CRs, with a simple averaging process, can achieve the same performance as the conventional hard decision CSS (soft decision CSS) using 40 CRs (30 CRs), respectively. This means that the proposed scheme succeeds to reduce the number of used CRs by 92.5 % (90 %) to obtain the same sensing performance like the conventional hard decision CSS (soft decision CSS), respectively. This can be considered as a significant complexity advantage because using a large number of CRs is not preferred in the CRN design. This is because the sensing time will be increased, which in turn results in reducing the overall CRN throughput. Beside reducing the total number of used CRs, the proposed scheme highly relaxes the requirement of high communication bandwidth required by the conventional soft decision CSS. Thus, using a lower communication bandwidth, the proposed hard decision CSS has a better sensing performance than the conventional soft decision CSS.

## VII. CONCLUSION

In this paper, an enhanced ED based hard decision CSS technique has been proposed to enhance the CRN SS performance. In the proposed scheme, the noise uncertainty in the CRs energy measurements is taken into account before arriving at a final decision about the current PU existence. Current PU status prediction is investigated by each CR using successive PU energy measurements during a specified duration. Based on this prediction, two dynamic thresholds are toggled to increase $P_d$ and decrease $P_{fa}$. The two thresholds are evaluated by each CR using an estimated value of its noise uncertainty factor. The enhanced false alarm and detection probabilities of the proposed scheme over the conventional hard decision CSS are theoretically proved. Besides, simulation results confirmed the theoretical claims and proved the high efficiency of the proposed scheme not only over the conventional hard decision based CSS but also over the conventional soft decision CSS using MRC, in which noise uncertainty effect is not considered. Moreover, the proposed scheme succeeded to reduce the number of used CRs by 92.5 % (90 %) to obtain the same performance like the conventional hard decision CSS (soft decision CSS), respectively.